\documentstyle[cite,cas,epsfig]{article}
\newcommand{\be}{\begin{equation}}
\newcommand{\ee}{\end{equation}}
\newcommand{\ba}{\begin{array}{c}}
\newcommand{\ea}{\end{array}}
\newcommand{\beqn}{\begin{eqnarray}}
\newcommand{\eeqn}{\end{eqnarray}}

\newcommand{\bi}{\begin{itemize}}
\newcommand{\ei}{\end{itemize}}

\newcommand{\cO}{{\cal O}}
\newcommand{\cP}{{\cal P}}

\newcommand{\rms}{\rm\scriptsize}
\def\tcf{$\tau$cF}
\def\jpsi{J/\psi}

\begin{document}

%
%

\begin{titlepage}
\vspace{0.3in}
\begin{flushright}
{CERN-TH.7066/93}
\end{flushright}
\vspace*{2.7cm}
\begin{center}
{\Large \bf PERSPECTIVES ON \\ TAU-CHARM FACTORY PHYSICS \\ }

\vspace*{0.4cm}
A. Pich$^{*\dagger}$

Theory Division, CERN, CH-1211 Geneva 23

\end{center}
\vspace*{1.5cm}

\centerline{ABSTRACT}

\noindent The Tau-Charm Factory combines the optimum conditions to perform
a high-precision investigation of the $\tau$ and $\nu_\tau$ leptons, the
$c$ quark,
and light-quark, glueball and hybrid spectroscopy.
An overview of the broad physics program that this facility
will address is presented.

\vspace*{2.2cm}
\begin{center}
{Physics Summary at the\\
Marbella Workshop on the Tau-Charm Factory \\
Marbella, Spain, June 1993}
\end{center}

\vspace*{2.5cm}

{\footnotesize\baselineskip 10pt \noindent
$^*$ On leave of absence from
Departament de F\'{\i}sica Te\`orica, Universitat de Val\`encia,
and IFIC, Centre Mixte Universitat de Val\`encia--CSIC,
E-46100 Burjassot, Val\`encia, Spain.}

{\footnotesize\baselineskip 10pt \noindent
$^\dagger$ Work supported in part
by CICYT (Spain), under grant No. AEN93-0234.}

\vfill
\begin{flushleft}
{CERN-TH.7066/93 \\
November 1993}
\end{flushleft}
\end{titlepage}

%

\title{Perspectives on Tau-Charm Factory Physics}
\author{A. Pich\thanks{On leave of absence from
Departament de F\'{\i}sica Te\`orica, Universitat de Val\`encia,
and IFIC, Centre Mixte Universitat de Val\`encia--CSIC,
E-46100 Burjassot, Val\`encia, Spain.}\ \thanks{Work supported in part
by CICYT (Spain), under grant No. AEN93-0234.}}
\institute{Theory Division, CERN, CH-1211 Geneva 23}
\maketitle \thispagestyle{plain}
\begin{abstract}
The Tau-Charm Factory combines the optimum conditions to perform
a high-precision investigation of the $\tau$ and $\nu_\tau$ leptons, the
$c$ quark,
and light-quark, glueball and hybrid spectroscopy.
An overview of the broad physics program that this facility
will address is presented.
\end{abstract}
\pagestyle{plain}

\section{INTRODUCTION}

Due to its enormous phenomenological success, the Standard Model (SM)
is now generally accepted as
the established theory of the electroweak interactions.
However, the SM leaves too many unanswered questions to be
considered as a complete description of the fundamental forces.
Clearly, New Physics has to exist.
Although we do not have at present a clear idea about the kind of dynamics
which could lie beyond the SM,
we do know the kind of problems it should address.
The mechanism responsible for the spontaneous symmetry breakdown (SSB) of
the $SU(2)\otimes U(1)$ symmetry is completely unknown.
The scalar sector of the SM provides a simple way of realizing
the SSB, but we do not have at present any experimental evidence of
its correctness.
Related to that, there is the problem of fermion-mass generation.
We know that there are three (and only three) families of
fundamental fermions,
and we have empirically learned the values of their masses and mixings.
We ignore, however, which mechanism generates the fermionic
mass matrices. Why the number of generations is just 3? Why the pattern
of masses and mixings is what it is?
Are the masses the only difference among the three families?
What is the origin of the flavour structure of the SM?
Which dynamics is responsible for the observed CP violation?

In order to get some hints
on all those problems, we need additional experimental information.
We should investigate the existence of the Higgs particle (or whatever
new object is playing the role of the SM Higgs) to learn about the
SSB mechanism. This obviously requires to build high-energy
machines to explore energies above the electroweak scale, i.e the
100 GeV to 1 TeV region.
The study of the flavour problem requires, however, a completely
different approach:
we need high-precision and high-luminosity
machines,
producing large amounts
of particles of a given flavour;
the so-called factories.

The light quarks and leptons are by far the best known ones.
Many experiments have analyzed in the past the properties of
$e$, $\mu$, $\nu_e$, $\nu_\mu$, $\pi$, $K$, $\ldots$
Moreover, future facilities like DA$\Phi$NE and KAON plan to
further improve this knowledge.
However, one na\"{\i}vely expects the heavy fermions
to be much more sensitive to
New Physics.
It is unfortunate that the present precision
in heavy-flavour experiments is only good enough to study the
gross properties of these particles.
Obviously, new facilities are needed
to  match (at least) the
precision attained for the light flavours.

Two different (and complementary) machines have been proposed to
perform a detailed study of heavy flavours: the B Factory (BF) and
the Tau-Charm Factory (\tcf).
Both  are high-luminosity
(${\cal L}\ge 10^{33} \mbox{cm}^{-2} \mbox{sec}^{-1}$)
$e^+e^-$ colliders, but with different centre-of-mass energies.
The BF would run in the $\Upsilon$ region, to produce a huge and
clean sample of
$B$--$\bar B$ pairs. It would be a dedicated machine for investigating
the properties of the $b$ quark, and  would provide a marvelous tool
to perform CP-violation studies.
The \tcf, on the other side, would run at lower energies, around the
$\tau^+\tau^-$ and $c\bar c$ thresholds,
in order
to make
a high-precision
analysis of the $\tau$, $\nu_\tau$  and $c$ in the absence of any
$b$ contamination.

It is worth stressing the special features of the three
poorly-known
fermions that  the \tcf\ would study:
\begin{itemize}

\item{The $\tau$ is a $3^{rd}$ generation lepton with a large number
of decay channels,
some of which are calculable with high-precision in the SM.
It is the heaviest known lepton and,
moreover, it is the only one heavy enough to decay into hadrons.
The pure leptonic or semileptonic character of $\tau$ decays
provides a clean laboratory to test the structure of the weak
currents and the universality of their couplings to the gauge bosons.}

\item{The $\nu_\tau$ is a quite unknown particle.
We have learn experimentally that the family partner of the $\tau$
is  different
from $\nu_\mu$ and $\nu_e$, and the present data is consistent
with the $\nu_\tau$ being a conventional sequential neutrino.
However, we are very far from clearly establishing its properties.}

\item{The $c$ is the only heavy up-type quark accessible to precision
experiments. It has a rich variety of weak decays (Cabibbo allowed,
Cabibbo forbidden, doubly Cabibbo forbidden, rare second-order decays,
\ldots), which can be used to learn about the interplay of weak
and strong interactions.
$D^0$--$\bar D^0$ mixing and CP-violation studies in the up sector
would be of enormous
interest. Moreover, the $c$ quark also provides --through decays
of the $J/\psi$, $\psi'$, \ldots -- an ideal tool for
light-meson and gluonium spectroscopy.}

\end{itemize}

The purpose of the \tcf\ is to measure the properties of these
particles, with precisions comparable to the ones
attained for the lighter fermions.
The \tcf\ would cover an exceptionally broad physics programme,
which contains three main elements:
\begin{enumerate}
\item Comprehensive precision tests of the electroweak parameters of the
SM.
\item Comprehensive precision tests of QCD at the interface of
perturbative and non-perturbative dynamics.
\item Search for new physics.
\end{enumerate}

At the very least, the first two items guarantee a substantial
contribution to our knowledge of the fundamental parameters
of the SM. The \tcf\ would ``write the book'' on $\tau$ and charm decays,
and spectroscopy. Moreover, it would be an ideal facility for making detailed
tests of non-perturbative QCD; the \tcf\ could well be considered
as ``the QCD machine of the next decade''.
In addition,
this program (item 3) may reveal clues to the origin of the family
puzzle or to physics beyond the SM.

The original ideas on the \tcf\ \cite{TCFa,TCFb} have been followed by
many detailed studies
\cite{tcws,studies,sevillaws}
on  the design
and feasibility of the \tcf\ detector and machine,
and the  physics programme that could be addressed with this facility.
Many new developments have been reported in this workshop.
In the following, I will  present a short
(and necessarily incomplete) update of the \tcf\ physics prospects.
I will focus on a few issues to illustrate
the physics potential of this machine.
Further details can be obtained
from the quoted references.

\section{ADVANTAGES OF THE THRESHOLD REGION}

The \tcf\ would generate a copious number of $\tau$'s and charm
particles, increasing the present data samples by two
or three orders of magnitude. However, as shown in
Table \ref{tab:data_samples}, this large statistics is not the main
advantage of the \tcf.
Although the highest cross-sections for $\tau^+\tau^-$ and $c\bar c$
occur near threshold,
other facilities, like the BF or an hypothetical
high-luminosity $Z$ factory, could produce a comparable
(slightly smaller) number of $\tau$'s and $D$'s (but not of
$J/\psi$'s).

\begin{table}[hbt]
\caption{Comparison of $\tau$-charm yearly data samples  at the
Z, B and $\tau$-charm factories. The quoted numbers correspond to
integrated luminosities of 2 fb$^{-1}$
(${\cal L}= 2\times 10^{32} \mbox{cm}^{-2} \mbox{s}^{-1}$)
for the Z factory and
10 fb$^{-1}$ (${\cal L} = 10^{33} \mbox{cm}^{-2} \mbox{s}^{-1}$)
for the BF and the \tcf.}
\label{tab:data_samples}
\begin{center}
\begin{tabular}{|l|c|c|l|}
\hline
Particle & Z Factory & B Factory & \ \ \ \ \tcf
\\ \hline
$D^0$ (single) & $1.2\times 10^7$ & $1.5\times 10^7$ &
    $5.8\times 10^7$  ($\psi''$)
\\
$D^+$ (single) & $0.5\times 10^7$ & $0.7\times 10^7$ &
    $4.2\times 10^7$  ($\psi''$)
\\
$D^+_s$ (single) & $0.3\times 10^7$ & $0.3\times 10^7$ &
    $1.8\times 10^7$  ($4.14$ GeV)
\\ \hline
$\tau^+\tau^-$ (pairs) & $0.3\times 10^7$ & $0.9\times 10^7$ &
 $\ba \!\!  0.5\times 10^7 \; \mbox{\rm (3.57 GeV)}
\\ \!\!
 2.4\times 10^7 \;  \mbox{\rm (3.67 GeV)}
\\ \!\!
 3.5\times 10^7 \; \mbox{\rm (4.25 GeV)}  \ea $
\\ \hline
$J/\psi$ & -- & -- & $1.7\times 10^{10}$
\\
$\psi'$ & -- & -- & $0.4\times 10^{10}$
\\ \hline
\end{tabular}
\end{center}
\end{table}

With the sharp increase in statistics foreseen in the future facilities,
the attainable precision will likely be limited by
backgrounds and systematic errors.
The decisive advantage
that makes the \tcf\ the best experimental tool for $\tau$
\cite{pich}
and charm \cite{roudeau}
physics,  is
its capability of tightly controlling the backgrounds and systematic errors.
The threshold region provides a unique environment, where backgrounds
are both small and experimentally measurable.
By adjusting the beam energy above or below a particular threshold,
the backgrounds can be directly measured, avoiding any need
for Monte Carlo simulations with their inevitable uncertainties.
Moreover, the data samples are very pure because they are free of
heavier flavour backgrounds. One can collect $\tau$ events that are
completely free of $c$ and $b$ contaminations, and charm samples are free
from $b$ backgrounds.

Near threshold, the heavy flavours are produced in simple
particle--antiparticle
final states
($\tau^+\tau^-$, $D^0 \bar D^0$, $D^+ D^-$, $\ldots$). By observing the decay
of one  particle, its partner is cleanly tagged. This ``single-tagging''
is a unique feature of the threshold region, which allows to collect
a bias-free data sample, without any pre-selection of its decay mode.
This property turns out to be crucial to achieve very precise measurements
of branching ratios. In addition, the threshold region has also several
kinematic advantages that result from the low particle velocities,
such as monochromatic spectra for two-body
decays.

  Another important advantage of the \tcf\ energy region is the existence
of two precisely-known energy points, the $J/\psi$ and $\psi'$, which
provide a very high-rate ($\sim 1$~kHz) signal to calibrate and
monitor the detector performance. This
allows for a tight control of the systematic errors.

Thus, the \tcf\ experiment would
benefit from a very
high statistics, low and measurable backgrounds, and reduced
systematic errors.
The coincidence of all these features near threshold
creates
an ideal facility for precision $\tau$-charm studies.

\begin{figure}[htb]
\caption{The cross-section ratio
$R\equiv \sigma(e^+e^-\to\mbox{hadrons},\tau^+\tau^-)/
\sigma(e^+e^-\to\mu^+\mu^-)$
in the $\tau$-charm threshold region.
The data are from DELCO \protect\cite{kirkby78} at SPEAR.}
\label{fig:delco}
\end{figure}

Figure \ref{fig:delco} shows the cross-section ratio
$R\equiv \sigma(e^+e^-\to\mbox{hadrons},\tau^+\tau^-)/
\sigma(e^+e^-\to\mu^+\mu^-)$
in the $\tau$-charm threshold region \cite{kirkby78}.
One can identify a series of
important operating energy points:
\begin{itemize}

\item{{\bf\boldmath $J/\psi$(3.10):}
Intense and clean source of light hadrons and
gluonic particles for QCD studies and tests of symmetry principles.
High-rate signal for calibration.}

\item{{\bf 3.55 GeV:}
Just below the $\tau^-\tau^+$ threshold.
Experimental measurement of all non-$\tau$ backgrounds.}

\item{{\bf 3.56 GeV:}
$\tau^+\tau^-$ threshold.
Due to the Coulomb
interaction, the  $\tau^+\tau^-$ production cross-section has a finite value
of 0.20 nb \cite{perrotet}.
The two body $\tau$ decays
($\tau\to\pi\nu_\tau, K\nu_\tau$) produce monochromatic secondaries.
Very clean signatures and good kinematic separation
of the different decay modes.}

\item{{\bf 3.67 GeV:}
The highest $\tau^+\tau^-$ cross-section below
the $\psi'$ and $D$--$\bar D$ thresholds.
$\tau$ decay is the only source of prompt single-charged leptons
and promt neutrinos.}

\item{{\bf\boldmath $\psi'$(3.69):}
Another high-rate source of light hadrons
for QCD studies and calibration.
The highest $\tau^+\tau^-$ cross-section occurs here; background-free
experiments, such as neutrinoless rare $\tau$-decays, could be made
at this energy.}

\item{{\bf\boldmath $\psi''$(3.77):}
Pure $D^0\bar D^0$ and $D^+D^-$ states,
without contamination from other charmed particles. Tagged $D$
decays.}

\item{{\bf 4.03 GeV:}
The highest charm cross-section in $e^+e^-$
annihilation. Tagged $D_s^\pm$ decays.}

\item{{\bf 4.14 GeV:}
$D_s^\pm$ studies, via $D_s^\pm D_s^{*\mp}$ events.}

\item{{\bf 4.57 (4.91, 4.93, \ldots) GeV:}
$\Lambda_c\;\bar\Lambda_c$
($\Sigma_c\;\bar\Sigma_c$, $\Xi_c\;\bar\Xi_c$, \ldots)
threshold.}

\end{itemize}

\section{TAU PHYSICS}

\subsection{Precise branching ratios}

  The leptonic or semileptonic nature of $\tau$ decays allows us
to make precise theoretical predictions for its decay rates and
Dalitz-plot distributions.
A systematic study of all $\tau$
decay channels at the \tcf\ could be used to look
for signs of discrepancies with the theoretical
expectations.

The simplest decay modes,
$\tau^-\to e^-\bar\nu_e\nu_\tau$, $\mu^-\bar\nu_\mu\nu_\tau$, $\pi^-\nu_\tau$
and $K^-\nu_\tau$,
are theoretically
understood  at the
level of the electroweak radiative corrections \cite{MS88}.
Within the SM, one can accurately predict the following relation
between the leptonic branching ratios
$B_l\equiv B(\tau^-\to\nu_\tau l^-\bar\nu_l)$ and the $\tau$ lifetime
\cite{TAUREV}:
\begin{equation}
\label{relation}
B_e \, = \, {B_\mu \over 0.97257} \, =
{ \tau_{\tau} \over (1.632 \pm 0.002) \times 10^{-12}\, {\rm s} } \, .
\end{equation}
To derive this equation, one uses the value of the Fermi coupling
measured in $\mu$ decay, and the recent precise measurement \cite{BES} of
the $\tau$ mass
[$\Gamma(\tau^-\to\nu_\tau l^-\bar\nu_l)\propto G_F^2 m_\tau^5$].
We can use this relation to test the universality of the
charged-current couplings to the $W$ boson.
Allowing these couplings to depend on the considered lepton flavour,
the ratio $B_\mu/B_e$ provides a measurement of $|g_\mu/g_e|$,
while $|g_\tau/g_\mu|$ can be obtained from the
ratio $B_e \tau_\mu/\tau_\tau$.
The present data \cite{fernandez,ohio} imply:
\begin{equation}
\label{eq:univI}
\Big\vert {g_{\tau} \over g_{\mu}} \Big\vert \, = \,
    0.995 \pm 0.007 \, ;
\qquad\qquad
\Big\vert {g_{\mu} \over g_e } \Big\vert \, = \,
   1.001 \pm 0.006 \, ,
\end{equation}
which should be compared with the more accurate value \cite{PION}
$\, |g_{\mu} / g_e | = 1.0014 \pm 0.0016 \, $
obtained from pion decay.

The decay modes $\tau^-\to\nu_\tau\pi^-$ and $\tau^-\to\nu_\tau K^-$ can also
be used to test universality,
provided the dependence on
the hadronic matrix elements (the so-called decay constants $f_{\pi,K}$)
is eliminated
through the ratios
$\Gamma(\tau^-\to\nu_\tau\pi^-/K^-)/\Gamma(\pi^-/K^-\to l^-\bar\nu_l)$.
The theoretical uncertainties from uncalculated higher-order electroweak
corrections have been conservatively estimated \cite{marciano92} to be about
$1\%$, but it seems feasible to reduce them to the $0.1\%$ level.
{}From the present data, one obtains:
\begin{equation}
\label{eq:univII}
\Big\vert {g_{\tau} \over g_{\mu}} \Big\vert  =
    1.007 \pm 0.013 \,  ; \quad\quad
\Big\vert {g_{\tau} \over g_{e}} \Big\vert  =
    1.010 \pm 0.013  .
\end{equation}
The formal average of Eqs. (\ref{eq:univI}) and (\ref{eq:univII}),
implies
$|{g_{\tau}/ g_{\mu}}| = 0.998\pm 0.006$,
in perfect agreement with the expected universality of the
charged-current couplings.

In the \tcf, the expected precisions on the 1-prong branching ratios are
0.1\% ($e,\mu,\pi$) -- 0.8\% ($K$)  in a one-year data sample
\cite{gomez},
whereas present errors are 1\% -- 30\%. These precise
measurements would allow a test of the universality of the leptonic
charged-currents at the 0.1\% level, to be compared with the present
accuracies of 0.7\%.
The experimental requirements to achieve
precise --${\cal O}(0.1\%)$-- measurements of the $\tau$ branching
ratios are very challenging; the normalization,  detection efficiency and
backgrounds must each be known at the 0.1\% level. These requirements can
probably be achieved only at the \tcf,  running just above
the $\tau^+\tau^-$
threshold at 3.56 GeV, where the decays are kinematically-separated.

The measured ratio between the
$\tau^-\to\nu_\tau K^-$ and $\tau^-\to\nu_\tau \pi^-$
decay widths can be used to obtain a value for
$\tan^2{\theta_c} (f_K / f_\pi)^2$,
\begin{equation}
\tan^2{\theta_c} \, \Bigl ( {f_K \over f_\pi} \Bigr )^2 \, =
    \, \Bigl ( {m^2_\tau - m^2_\pi \over m^2_\tau - m^2_K } \Bigr )^2
    \, { \Gamma (\tau^-\to\nu_\tau K^- ) \over \Gamma (\tau^-\to\nu_\tau
    \pi^- ) } \, = \, (7.0 \pm 2.0) \times 10^{-2} .
\end{equation}
The present accuracy (30\%) is very poor
(a better 12\% precision is quoted in preliminary LEP results
\cite{fernandez}).
The excellent $\pi/ K$ separation of the \tcf\ will make possible
to reach a precision of $0.8\%$ (0.5\%) in a one-year (two-year) data sample,
which is comparable to the result
$(7.56 \pm 0.02) \times 10^{-2}$,
obtained from  \cite{PDG92}
$\Gamma (K^- \to \mu^-\bar\nu_\mu ) / \Gamma (\pi^-\to\mu^-\bar\nu_\mu )$.

\subsection{Hadronic $\tau$ decays}

   The semileptonic decay modes of the tau, $\tau^-\to\nu_\tau H^-$,
probe the  matrix
element of the left-handed charged current between the vacuum and the
final hadronic state $H^-$.
%
%
Contrary to the well-known \ $e^+e^-\to\gamma\to$ hadrons \  process,
which only tests the electromagnetic vector current, the semileptonic
$\tau$-decay modes offer the possibility to study the properties of both
vector and axial-vector currents.

Since the  hadronic matrix elements are governed by the non-perturbative
regime of QCD, we are unable at present to make first-principle
calculations for exclusive decays.
Nevertheless, we can use our present knowledge on
strong interactions at low energies to estimate the gross features
of the Lorentz-invariant form factors describing the
decay amplitudes.
For instance, Chiral Perturbation Theory techniques \cite{mexico}
can be applied to rigorously predict the low hadronic-invariant-mass
behaviour, and one can extrapolate to higher values of $q^2$
by using different models of resonance dynamics
\cite{FWW80,pichslac,pich87,JJCONPICH}.

The hadronic form factors can be experimentally extracted from the
Dalitz-plot distributions of exclusive hadronic $\tau$ decays \cite{mirkes}.
An exhaustive analysis of those decay modes at the \tcf\
would provide a very valuable data basis to confront with
theoretical models.
The high statistics of the \tcf\ would even make possible
to perform a study of the $J=0$ channels, which are very sensitive to the
masses of the light quarks;
the accurate measurement of certain azimuthal asymmetries
could then result in an experimental determination of the
running quark masses \cite{stern}.

At the inclusive level, the present QCD techniques are much more powerful.
The total inclusive
hadronic width of the $\tau$ can be systematically
calculated \cite{BNP,BRAATENa}  by using analyticity constraints and the
Operator Product Expansion.  Both perturbative (to order $\alpha_s^3$) and
non-perturbative (which have been shown to be suppressed) contributions
have been taken into account \cite{BNP,BRAATENa},
together with the known electroweak corrections \cite{MS88,BRAATENb}.
The result turns out to be quite sensitive to the value of $\alpha_s$,
and has been used to obtain one of the most precise
determinations of the strong coupling constant
\be
\alpha_s(m_{\tau}) = 0.35 \pm 0.03 .
\ee
After evolution
up to the scale $M_Z$, this running coupling constant decreases to
$\alpha_s(M_Z)  =  0.121\pm 0.004$,
in amazing agreement with the
present LEP average \cite{LEP} (without $\tau$ data)
$\alpha_s(M_Z) \, = \, 0.123\pm 0.005$
and with a similar error bar.
The comparison of these two determinations of $\alpha_s(\mu)$ in two extreme
energy regimes, $m_\tau$ and $M_Z$, provides a beautiful test
of the predicted running of the QCD coupling constant.

The non-strange and strange components of the $\tau$ hadronic width
can also be predicted separately, and
one can decompose these further into vector and axial-vector
contributions, which are resolved experimentally according to whether
the hadronic final state includes an even or odd number of pions.
A consistency check of the theoretical analysis can be done by studying
the distribution of the final hadrons in $\tau$ decay. A
combined fit of certain weighted integrals \cite{LDPb,diberder}
 of the hadronic spectrum would allow
to simultaneously measure
$\alpha_s(m_\tau)$ and the parameters characterizing the non-perturbative
dynamics.
Moreover,
the measurement of the hadronic spectrum could provide
a precise determination of
the vector- and axial-vector-current spectral functions, both in the
Cabibbo-allowed and Cabibbo-suppressed channels.
This information would be very valuable for making many interesting tests
of QCD \cite{pichslac}.

A pioneering QCD analysis of the tau data has recently been
performed by the ALEPH collaboration \cite{diberder,ALEPH93}.
Their results are in good agreement with the theoretical expectations;
however, the present precision is still not good enough
for making a detailed test of the QCD predictions.
The \tcf\ is ideally suited for this kind of physics \cite{diberder}:
high statistics, low backgrounds, excellent momentum resolution,
easy $\pi / K$ identification, high resolution for both charged and
neutral measurements, and zero normalization uncertainty (single tagging).
The \tcf\ could perform a precision
global analysis of all $\tau$ decay channels,
with notable strengths for decays involving $K$'s and multiple $\gamma$'s,
which are poorly measured at present.
This should definitively settle present experimental controversies,
such as the so-called
1-prong problem, and would
provide a beautiful scenario to test QCD at low energies.

\subsection{Rare and forbidden decays}
\label{sec:rarebrs}

In the minimal  SM with massless neutrinos, there is a
separately conserved additive lepton number for each generation. All
present data are
consistent with this conservation law. However, there are no strong
theoretical
reasons forbidding a mixing among the  different  leptons, in the same
way as happens in the quark sector.
Many models, in fact, predict lepton-flavour or even
lepton-number violation at some level.
Experimental searches for these
processes
can provide information on the scale at which the new physics begins to
play a  significant role.

$K$, $\pi$ and $\mu$ decays, together with $\mu$--$e$ conversion,
neutrinoless double beta
decays  and  neutrino  oscillation  studies,  have  put  already
stringent  limits  \cite{PDG92} on
lepton-flavour and lepton-number violating interactions.
However, given the present lack of
understanding of the origin of fermion generations, one can imagine
different
patterns of violation of this conservation  law  for  different  mass
  scales.
For instance,
since the $\tau$ is a third generation lepton, new
physics associated with Higgs  couplings is expected to give larger effects
than for the lighter leptons.
Moreover, the larger mass of the tau lepton
also means that many more decays are kinematically allowed.
Several supersymmetric models lead to lepton-number
violating rates which may be constrained at the
\tcf.
These type of processes also appear \cite{concha} in models  with
extra neutral isosinglet or vector-like heavy fermions.

The present upper limits on lepton-flavour and
lepton-number violating decays of the tau
are in the range of $10^{-4}$ to $10^{-5}$, which is far
away from the   impressive   bounds  \cite{PDG92}
obtained   in $\mu$-decay.
The \tcf\ could easily improve these limits by about two
orders of magnitude.

\begin{figure}[htb]
\caption{The combined electron spectra (solid histograms) from the standard
decay  $\tau^-\to e^-\bar\nu_e\nu_\tau$
mixed with a hypothetical 1\%
two-body decay $\tau\to eX$  at   centre-of-mass energies
of a) 3.57 GeV  and b) 10 GeV \protect\cite{concha}.
The particle $X$ is a
massless Goldstone boson.  The dashed histograms show conventional (V-A)
fits to the combined spectra.  Each plot contains 200k
$\tau^-\to e^-\bar\nu_e\nu_\tau$,
corresponding to two months' data at 3.57 GeV. }
\label{fig:pex}
\end{figure}

The \tcf\ is especially suited to
study two-body decays like $\tau^-\to l^- X$
($l=e,\mu$; $X$ = Majoron, familon, flavon, \ldots), which,
at threshold,
lead to the distinctive signature of monochromatic leptons (Fig.
\ref{fig:pex}a)\cite{concha}.  In contrast, the sensitivity is much weaker at
higher energies since the lepton spectrum from the $l^- X$ decay is
broad,  spreading over the full spectrum of the
standard $l^-\bar{\nu}_l\nu_{\tau}$ decay (Fig. \ref{fig:pex}b).
For $\tau^-\to e^- X$ ($\tau^-\to \mu^- X$), the
expected branching-ratio sensitivity at the \tcf\  is better than $10^{-5}$
($10^{-3}$)
for one-year's data \cite{concha}, to
be compared with the present limits of 1-2\%.  The experimental sensitivity
could be improved a further order-of-magnitude if the monochromator
optics is successful.  Sensitivity to still-lower
branching ratios would require improvements in the $\pi e$ rejection of the
\tcf\ detector, e.g. with a fast RICH.

The optimum energy to search for other neutrinoless $\tau$ decays is
at 3.67 GeV, where a simple and ($\simeq$~100\%) efficient tag,
$E_{\mbox{\rms miss}}\geq$
0.8 GeV, can be used.  The validity of any signal can be convincingly
demonstrated by its disappearance at 3.55 GeV.   The \tcf\ will be able to
place upper limits of Br $\simeq 10^{-7}$ on these decays
\cite{concha}, or observe them at
a higher value.
If the  monochromator
optics is successful, it may be possible to reach even higher sensitivities for
certain  rare $\tau$ decays at the $\psi'(3.69)$.
Here the production rate is at a
maximum  and the backgrounds for neutrinoless
(fully-reconstructed) $\tau$ decays are small.

The \tcf\ would also look for $\tau$ decays which, although allowed in
the SM, are highly suppressed. For instance,
the decay $\tau^- \rightarrow \pi^-\eta\nu_\tau$ can only  occur through
second-class weak currents (which have never been observed) or through
non-standard mechanisms such as the violation of G parity or the existence
of a new scalar particle. The branching ratio expected in the
SM due to the violation of G parity (from the $u-d$ quark mass
difference) is about $10^{-5}$\cite{pich87}.   A Monte Carlo  study shows that
this decay  can be cleanly measured at 3.67 GeV \cite{gan}.

\subsection{Lorentz structure of the charged weak current}
\label{sec:current}

The V$-$A structure of the charged current can be tested by studying the
distribution of the final charged lepton in the leptonic decay modes of
the $\tau$,
$\, \tau^-\to\nu_{\tau} l^- \bar\nu_l \, $ ($l = e, \mu $).
This distribution is usually parametrized in terms of the
so-called Michel parameters,
which have the following SM values:
$\rho = \delta = 3/4 , \eta = 0 $ and $\xi = \xi' = 1 $.

Up to now, the experiments were only sensitive to the
parameter $\rho$.
The first measurement of $\xi$ has been recently reported \cite{ARGUSxi}
by ARGUS.
Averaging the different experiments one has \cite{TAUREV,ARGUSxi}:
\beqn
\label{eq:rho}
\rho_e & = &  0.717\pm 0.038 \,  , \quad
          \rho_{\mu} \, = \, 0.762\pm 0.046\, , \\
 \xi &\equiv &\sqrt{\xi_e\xi_\mu} \, = \, 0.90\pm 0.15\pm 0.10 .
\eeqn
The present measurements are in agreement with the V$-$A hypothesis.
Assuming that the $\tau \nu_{\tau} W$ vertex is a linear
combination of vector and axial currents,
$g_V V^\mu + g_A A^\mu$,
and using the standard V$-$A form for the
$l^- \nu_l W$ ($l = e, \mu $) vertex,
one can exclude a vertex
of pure V+A, V or A type. However, one can only obtain \cite{TAUREV}
the  upper limit
$|(g_V + g_A)/(g_V - g_A)|<0.33$ (95\% C.L.)
on a possible mixture of right-handed current structure; a very
poor limit indeed.

 The  most
general,
local, derivative-free lepton-number conserving four-fermion
interaction Hamiltonian   ($l = e, \mu $),
\be
{\cal H} \, = \, 4 {G_0 \over \sqrt{2}} \, \sum_{n,
    \epsilon, \omega} \, g^n_{\epsilon\omega} \,
    [\bar l_\epsilon \Gamma^n (\nu_l)_{\sigma} ] \,
    [(\bar\nu_\tau)_{\lambda} \Gamma_n \tau_\omega ] \, ,
\ee
contains ten complex coupling constants or, since a common phase is
arbitrary, nineteen independent real parameters.
$\epsilon , \omega , \sigma, \lambda$ are the chiralities (left-handed,
right-handed)  of the  corresponding  fermions, and $n$ labels the
type of
interaction (scalar, vector, tensor); for given $n, \epsilon ,
\omega $, the neutrino chiralities $\sigma $ and $\lambda$
are uniquely determined.
In the SM, $g^V_{LL}  = 1$  and all other
$g^n_{\epsilon\omega} = 0 $.
The normalization is conventionally fixed by
taking out a common factor $G_0$, which is determined by the total
decay rate.
With the present experimental information, it is certainly impossible
to determine the interaction.
For instance,
any combination of the six couplings
$g^V_{LL} , g^V_{RR} , g^S_{LL} , g^S_{RR} , g^S_{LR}$
and $g^S_{RL}$, with the other four couplings being zero, yields
$\rho = 3/4 $ \cite{FETSCHER}.

For the analogous $\, \mu^- \to\nu_\mu e^- \bar\nu_e \, $   decay,
Fetscher et al.  \cite{fgj86} have succeed in
proving the V$-$A nature of the $\mu$-decay amplitude,
using existing data. The relevant experiments
involve measurements of normal $\mu$ decay,
including the decay asymmetry relative to the spin of the muon
($\xi , \delta $) and the polarization of the final electron ($\xi'$),
together with the cross-section for the inverse $\mu$ decay
$ \, \nu_\mu e^- \to \mu^- \nu_e \, $ and the
value of the $\nu_\mu$-helicity,
which is known from the decay of the parent pion.
It would be desirable to do a similar analysis for the leptonic
tau  decay  modes.

Low energy experiments near the $\tau^+ \tau^- $ threshold are best
suited for more precise determinations of the $\rho$ parameter,
because the energy spectrum is less
distorted by the Lorentz boost of the moving $\tau$ .
The   measurement   of   the
low-energy parameter $\eta$ seems
possible for the $\, \tau^-\to\nu_\tau \mu^- \bar\nu_\mu \, $
mode, where the mass suppression is weaker
($m_\mu / m_\tau \approx 1 / 17 $).
Since
$\Gamma_{\tau^-\to\nu_\tau l^-\bar\nu_l}\sim (1 + 4 \eta_l m_l/ m_\tau )$,
the value of $\eta_\mu$ is needed in order to fix the global strength
of the interaction, i.e. to derive the Fermi coupling constant and make
a test of universality \cite{FETSCHER}.
$\eta \not= 0 \, $ would show
that there are at least two different
couplings with opposite chiralities for the charged leptons.
If we assume the V$-$A coupling $g^V_{LL}$ to be dominant, then the second
coupling would be a Higgs type coupling $g^S_{RR}$ with right-handed
$\tau$ and $\mu$ \cite{FETSCHER}.

Measuring the $\tau$ polarization, two  more  decay  parameters,
 $\xi$ and $\delta$, can  be
determined.
At LEP $\tau$'s are polarized, because of the $Z$ coupling, which makes
the measurement feasible \cite{kounine}; however, one actually
measures the product $\xi\cP_\tau$. Thus,
$\xi=1$ is usually assumed,
to get a determination of the neutral-current
couplings via ${\cal P}_\tau$. In order to measure the $\tau$-decay
parameters
at LEP, one either needs to fix the neutral-current couplings
\cite{privitera}
from other  observables not related to ${\cal P}_\tau$
or perform a correlated analysis of the two $\tau$'s
\cite{correlations}.

At the \tcf\ the $\tau$'s are produced unpolarized. Nevertheless,
the fact that the spins
of the two $\tau$'s
are strongly correlated \cite{correlations,TSAI}
makes the experiment feasible.
One considers the process
$ \, e^- e^+ \, \to \,  \tau^- \tau^+
   \, \to \, (X_1^- \nu_\tau ...) \, (X_2^+ \bar\nu_\tau ...)$,
and computes the cross-section for the production of the particles
$X_1$ and $X_2$,
performing  a coherent  sum  over  the  (unobserved)  spins  of  the
$\tau$'s.  The correlated distribution of the final products $X_1^-$ and
$X_2^+$ carries
information on the spin-dependent part of the production amplitude.

 Assuming an ideal
detector with no systematic uncertainties and 100\% efficiency,
the expected sensitivity \cite{gomez,FETSCHER,stahl}
on the Michel parameters for $10^7$
$\tau^+\tau^-$ pairs is indicated in Table~\ref{tab:michel}.
These numbers only
take into account  the information obtained
from correlated events where both $\tau$'s decay into leptons.
The estimated error on $\rho$ is about the same as the present error
in $\mu$ decay, and $\eta_\mu$ could be measured  a factor of 5
more precisely than in $\mu^-\to \nu_\mu e^-\bar\nu_e$.
$\xi$ and $\delta$ could be measured with
slightly larger
errors than in $\mu$ decay.
Since the values of $\xi$ and $\delta$ are strongly anticorrelated,
the combination
${1\over 2}\{1 + {1\over 9}(3\xi-16\xi\delta)\}$,
which gives the probability that the decaying  $\tau$ was right-handed,
can be
more precisely measured, with an expected accuracy of 2\%.

\begin{table}[htb]
\caption{Expected errors \protect\cite{stahl}
on the Michel parameters for $10^7$ $\tau^+\tau^-$
pairs produced at $\protect\sqrt{s} = 4$ GeV, from energy--energy--angle
correlations in pure leptonic decays.
Better precisions may be reached if hadronic decays are included
(see text).
For comparison the present world averages \protect\cite{PDG92}
in $\mu$ decay are shown in the
last column.}
\label{tab:michel}
\begin{center}
\begin{tabular}{|c|c|c|c|}
\hline
& $\tau^-\to\nu_\tau e^-\bar\nu_e$ &
$\tau^-\to\nu_\tau \mu^-\bar\nu_\mu$ &
$\mu^-\to\nu_\mu e^-\bar\nu_e$
\\ \hline
$\rho$ & $\pm 0.0022$ & $\pm 0.0023$ & $\phantom{-}0.7518\pm0.0026$
\\
$\eta$ & $\pm 0.6200$ & $\pm 0.0035$ & $-0.0070\pm 0.0130$
\\
$\xi$ & $\pm 0.0350$ & $\pm 0.0350$ & $\phantom{-}1.0030\pm0.0080$
\\
$\delta$ & $\pm 0.0270$ & $\pm 0.0280$ & $\phantom{-}0.7490\pm0.0040$
\\ \hline
\end{tabular}
\end{center}
\end{table}

The next step would be to measure the polarization of the charged
lepton emitted in the $\tau$-decay.
This could be possible, in principle, for the decay
$\, \tau^-\to\mu^-\bar\nu_\mu\nu_\tau \, $
by stopping the muons and detecting their decay products.
A precision of about $15\%$ in the corresponding decay parameter
$\xi_\mu'$ could be reached \cite{FETSCHER}.

Including the correlations of the leptonic decays with the
hadronic ones,
the number of useful $\tau$ pairs would increase by more than a factor
of 6.
Moreover, the lepton--hadron correlations are  more sensitive
to some Michel parameters than the lepton--lepton ones,
because of the high spin-analyzing power of the hadronic two-body
decays.
Therefore, the accuracy could be improved by a substantial amount.
Preliminary analyses \cite{stahl2} suggest
that it could be possible to reach sensitivities
as good as
$\pm 0.0002$ ($\rho$), 0.1 ($|\eta_e|$), 0.001 ($|\eta_\mu|$),
$\pm 0.002$ ($\xi$) and 0.0003 ($\delta\times\xi$). Clearly,
more detailed studies are needed in order to get a reliable
estimate of the ultimate \tcf\ precision, but the prospects
look very encouraging.

Assuming in the hadronic channels
an arbitrary combination of vector and axial
couplings for the $\tau\nu_\tau W$ vertex,
the neutrino helicity
\be
h_{\nu_\tau} \equiv  {2 g_V g_A \over |g_V|^2 + |g_A|^2} ,
\ee
which plays the same role as $\xi$ in  leptonic decays,
could be measured to 0.3\% accuracy \cite{FETSCHER,stahl};
to be compared with the present value \cite{ARGUSMICHEL}
$h_{\nu_\tau}  = -1.01 \pm 0.08$.

None of these estimated accuracies takes efficiencies and
systematic errors
into account. While it is clear that a BF, with a similar number
of $\tau$ pairs,
could reach a comparable statistical precision
\cite{FETSCHER,weinstein},
only at the \tcf\ the assumption of negligible
systematic errors and 100\% efficiency can be justified \cite{stahl}.

\subsection{$\nu_\tau$ mass}

The possibility of a non-zero neutrino mass is obviously a
very important
question in particle physics. There is no fundamental principle requiring
a null
mass for the neutrino. On the contrary, many extensions of the SM
predict non-vanishing neutrino masses, which could have, in addition,
important implications in cosmology and astrophysics.

The first attempts to place a limit on $m_{\nu_\tau}$
were done by studying the endpoint of the
momentum spectrum of charged leptons from the decays
$\tau^-\to\nu_\tau l^- \bar\nu_l \quad (l = e,\mu)$. The
precision
which can be achieved is limited by the experimental momentum resolution
of fastest  particles,  which  deteriorates  with  increasing
centre-of-mass energy.
Better limits have been set by studying the endpoint of the
 hadronic
mass spectrum of high multiplicity tau decays.
The limiting factor is then the resolution of the effective
hadronic-mass determination. The strongest bound up to date
\cite{NEUTRINO} comes from
the $\, \tau^-\to 2 \pi^+ 3 \pi^- \nu_\tau \, $ decay,
\be\label{eq:numasslimit}
m_{\nu_\tau} \, < \, 31 \,\mbox{\rm MeV} \quad (95\%\,
\mbox{\rm C.L.}).
\ee

In order to substantially improve  the present
$m_{\nu_\tau}$  upper bound,
it is extremely important   to obtain a very clean
event sample, since a single background event near the endpoint can falsely
lower the final limit \cite{cowen}.
The advantages of the \tcf\ for this experiment are then obvious.
It has been shown that the study of
the decays $\tau^-\to\nu_\tau l^- \bar\nu_l$
could result in a limit of the order of 20 MeV \cite{gomez,JJCONCHA},
thus providing little improvement of the current limit.
The prospects are much better for the
hadronic  modes $\tau^-\to 2 \pi^+ 3 \pi^- \nu_\tau$ \cite{JJCONCHAII}
and $\tau^-\to K^- K^+ \pi^- \nu_\tau$ \cite{JJCONPICH},
where mass limits of about 5 and 10 MeV, respectively,
could  be easily
achieved, in a one-year data sample.
The limit can be further improved using the decay
$\tau^-\to 2 \pi^- \pi^+ 2\pi^0 \nu_\tau$; owing to its
high branching ratio, this mode has been shown to be very efficient
in a recent CLEO analysis \cite{cowen,CLEO}.
The very good photon resolution of the \tcf\ detector would guarantee
an optimal use of decays containing neutrals.
Adding the information from the the three decay modes, the
estimated sensitivity \cite{gomez} of the \tcf\ in a two-year data sample
is 2 MeV (95\% C.L.).

For comparison, the best limits on the muon and electron neutrinos are
\cite{PDG92} $m_{\nu_\mu} < 270$ KeV (90\% C.L.) and $m_{\nu_e} < 7.3$ eV
(90\% C.L.). Note, however, that
in many models a mass hierarchy among
different generations is expected, with the neutrino mass being
proportional to
some power of the mass of its charged lepton partner.
Assuming for
instance the  fashionable  relation
$\, m_{\nu_\tau} / m_{\nu_e} \sim (m_\tau/m_e)^2$,
a sensitivity of 2 MeV for $m_{\nu_\tau}$ is equivalent to 0.2 eV for
$m_{\nu_e}$.
 A
relatively crude measurement of $m_{\nu_\tau}$
may then imply strong constraints on neutrino-mass model building.

\subsection{$\tau$ mass}

  The mass is a basic property of any particle, which obviously
should be known as accurately  as possible.
A precise value of $m_\tau$ is needed in order
to improve the present bound on  $m_{\nu_\tau}$, or
to make a test of universality.
The recent accurate measurement of $m_\tau$ by the BEPC/BES collaboration
\cite{BES},
7 MeV lower than the old value quoted by the particle data group
\cite{PDG92},
has shown the relevance of knowing the exact value of this parameter.
Given the important implications of this measurement,
a confirmation of the $m_\tau$ value
to a similar (or better) level of precision
by an independent experiment is clearly called for.
This can only be done near threshold, by
measuring $\sigma(e^+e^-\to\tau^+\tau^-)$.
The \tcf\ could get a precision of 0.1 MeV on $m_\tau$,
which would represent a factor of 3 improvement over the BES measurement.

\subsection{Electromagnetic and weak moments}

A precision analysis of the $\tau^+\tau^-$ production cross-section
near threshold (expected accuracy at the \tcf\ $\sim$ 0.1\%)
could substantially improve the bound on the $\tau$
anomalous magnetic moment, reaching a sensitivity
\cite{g-2}
at the level of the first QED contribution ($\alpha/2\pi$).
This quantity is very poorly known
at present:
$a^\gamma_\tau < 0.11$ \cite{GM91}.

Non-zero electric dipole moments of leptons, $d_l$, are
unambiguous signals of T (CP) violation, and are sensitive
probes of physics beyond the SM.
The present limit on the $\tau$ electric dipole moment could be improved
by more than one order of magnitude by studying T-odd triple
correlations of the final $\tau^+\tau^-$ decay products.
A conservative estimate of the \tcf\ capabilities gave as
expected sensitivity $|d_\tau| \le 10^{-17}$ e cm \cite{nachtmann}.
As demonstrated by the recent LEP analyses of the analogous weak dipole
moment \cite{kounine}, a much better precision can probably be achieved.


\section{CHARM PHYSICS}

\subsection{Semileptonic decays and CKM matrix elements}

Semileptonic decays of heavy mesons play a crucial role in the
determination of the Cabibbo--Kobayashi--Maskawa (CKM) mixing
matrix.
The partial decay width associated with the decay
$M_i\to M_f l\nu_l$ is given by the product
$\Gamma = |V_{Qq}|^2 \widetilde{\Gamma}$,
where $V_{Qq}$ is the relevant quark-mixing
factor, and
the dynamical information is encoded in the
quantity $\widetilde{\Gamma}$, which depends
on the appropriate hadronic matrix element (squared)
of the left-handed current
(integrated over phase space with trivial leptonic and kinematical factors).
Unfortunately, $\widetilde{\Gamma}$
is governed by non-perturbative hadron dynamics
and, therefore, it is very difficult to give a theoretical prediction from
first principles.
One needs to rely in model-dependent estimates, usually combined
with approximate symmetry properties which can (in some cases) fix
the hadronic form factors in one point of the phase space,
providing  the needed normalization.
To determine the CKM mixing factor,
it is then crucial to have very good  data on the
distribution of the final decay products, in order to
experimentally measure the shape of the hadronic form factors.

There is a dual motivation to perform an accurate analysis of
semileptonic charm-decay modes at the \tcf:
\begin{itemize}
\item The determination of $V_{ub}$ (very poorly known at present) is
one of the main purposes of the BF.
Any model  that makes predictions about $B$-decay matrix elements
can be tested in $D$ decays, where non-perturbative effects are
magnified.
Once the non-perturbative corrections are understood in charm decays,
they can be scaled to beauty decays with a considerable degree of
confidence.
\item Although constrained by SM unitarity to 0.1\% ($V_{cs}$)
and 1\% ($V_{cd}$), the CKM charm matrix elements are poorly
measured at present ($\pm\: 10-20\% $).
A direct experimental determination with much better precision should
be done.
\end{itemize}

 At the \tcf\ the $D$ semileptonic
branching ratios could be measured with an accuracy better than 1\%
\cite{roudeau,izen,schindler},
whereas present errors are 9\% for $D^0\to K^-
e^+\nu_e$ and 50\% for $D^0\to \pi^- e^+ \nu_e$.
By systematically measuring the decay rates
and the corresponding
invariant-mass distributions of the final hadrons,
both in $D_{l3}$ and $D_{l4}$ decays,
the \tcf\ could determine with high precision
the shape and relative normalization of
all relevant
axial and vector form-factors,
providing the necessary information
to select among or improve the existing models
\cite{bigi,pene,form_factors}
of $D$ and $B$ meson decays.
A much precise and reliable extraction of CKM matrix elements would then
become possible.

Theoretical uncertainties are largely avoided by taking ratios,
such as  $\Gamma(D\to\pi l\nu)/\Gamma(D\to K l\nu)$,
where the form-factor uncertainty is reduced to the level of $SU(3)$
breaking.
 In this way, $V_{cd}/V_{cs}$ could be determined to 1\%,
which is comparable to the present precision on $\theta_{\mbox{\rms Cabibbo}}$.

The \tcf\ could   analyze in detail
the semileptonic decays of $D^0$, $D^+$, $D_s^+$,
$\Lambda_c$ and $\Xi_c$, providing an exhaustive data basis
to confront with the predictions \cite{bigi,pene,form_factors}
of present theoretical technologies
(heavy-quark expansions, QCD sum rules, lattice, \ldots).
A detailed study of the inclusive semileptonic $c\to s$ and $c\to d$
decays would be of particular value \cite{bigi}.

\subsection{Leptonic $D$ decays}
\label{sec:leptonic}

Pure leptonic decays of $D^\pm_{(s)}$ mesons  can be
rigorously calculated in the SM. The predicted branching
ratios are
\begin{equation}
\label{eq:dleptonic}
         Br( D^+_{(s)} \rightarrow   l^+\nu_l ) =
   \tau_{D^+_{(s)}} {G_F^2  \over  8 \pi}
  f_{D_{(s)}}^2  m_{D^+_{(s)}} \mid V_{cd(s)} \mid^2  \times
  m_l^2   \left( 1 - { m_l^2      \over m_{D^+_{(s)}}^2  } \right)^2  ,
\label{eq:leptonic1}
\end{equation}
where $f_{D_{(s)}}$ is the so-called weak decay constant,
which measures the wave-function overlap of
the $c$ and $d(s)$ quarks in the $D^\pm_{(s)}$ meson.

The meson decay constants are used to predict non-leptonic decays and
second-order weak processes --including mixing and
CP violation--, and are therefore important quantities to be experimentally
determined. Precise measurements of these constants would also provide
important tests of  modern methods of calculation in non-perturbative
QCD \cite{martinelli,dominguez,shifman},
thereby providing a firmer
ground for attemps to extrapolate to beauty systems (since $f_B$  is
experimentally inaccessible) and allowing a better theoretical
description of $D^0$--$\bar D^0$ mixing.

In the SM,
leptonic $D$ decays are helicity suppressed,
and therefore the
decay amplitudes are proportional to the final charged-lepton mass.
Taking $f_D\sim 170$ MeV and $f_{D_s}\sim 200$ MeV
\cite{martinelli,dominguez},
the largest pure leptonic branching ratios
are estimated to be $Br( D^+_s
\rightarrow   \tau^+\nu_\tau ) \simeq 3$\%, $ Br( D^+_s \rightarrow
\mu^+\nu_\mu ) \simeq 3 \times 10^{-3}$, and $ Br( D^+ \rightarrow
\mu^+\nu_\mu ) \simeq 3 \times 10^{-4}$.
This should be compared with
the upper bounds quoted by the particle data group \cite{PDG92}:
$ Br( D^+_s \rightarrow
\mu^+\nu_\mu ) < 3\% $, and $ Br( D^+ \rightarrow
\mu^+\nu_\mu ) < 7.2 \times 10^{-4}$
(preliminary evidence of leptonic $D_s$ decays has been reported
recently \cite{witherell} by several experiments).

The theoretical uncertainties associated with the meson decay constants
are much smaller for the ratio $f_D/f_{D_s}$, which is only sensitive to
$SU(3)$ breaking.
A precise measurement of the ratio
$Br( D^+ \rightarrow   \mu^+\nu_\mu )/
Br( D^+_s \rightarrow   \mu^+\nu_\mu )$
would allow to make an independent  determination of
$|V_{cd}|/|V_{cs}|$.

Furthermore, the relative value of the $D_s^\pm$ leptonic branching ratios
is
precisely predicted in the SM:
\be\label{eq:ds_leptonic}
{Br( D^+_s \rightarrow   \mu^+\nu_\mu ) \over
           Br( D^+_s \rightarrow   \tau^+\nu_\tau )}   =
{m_\mu^2   \left[ 1 - ( m_\mu^2    /     m_{D_s}^2  ) \right]^2  \over
 m_\tau^2   \left[ 1 - ( m_\tau^2   /     m_{D_s}^2  ) \right]^2} =
 0.102 \pm 0.001 .
\ee
An accurate
measurement of the pure leptonic braching ratios
would then provide a novel test of $\mu-\tau$ universality, involving
the process:
 \[    2^{\mbox{\rms nd}} \; \mbox{\rm family quarks} \;
\stackrel{W,X?}{\longrightarrow} \; \left\{
\begin{array}{c}
 2^{\mbox{\rms nd}} \; \mbox{\rm family  leptons}  \\
\mbox{\rm vs.}  \\
3^{\mbox{\rms rd}} \; \mbox{\rm family  leptons}  \\
\end{array}
\right. .  \]
The ratio (\ref{eq:ds_leptonic}) is
sensitive to new physics ($X$) that does not have
the usual helicity suppression of pseudoscalar-meson decay or that
has mass-dependent couplings.               The absence of
$e\nu$ final states would provide a further test of the SM prediction.

Precise measurements of
 pure leptonic $D$ decays require single-tagged event samples
and are therefore  uniquely accessible at the \tcf. Tagged event samples
are necessary both to suppress backgrounds and to provide a constrained fit
for the mass of the missing $\nu$('s).
A detailed Monte
Carlo study \cite{schindler,kim} indicates that the signals should be clearly
distinguished from background processes. With 1-year's
data, each of the the branching ratios
$ Br( D^+_s \rightarrow   \tau^+\nu_\tau )$,
   $ Br ( D^+_s \rightarrow   \mu^+\nu_\mu )$  and
 $ Br ( D^+ \rightarrow   \mu^+\nu_\mu )$ could  be measured to about 2\%
accuracy.

\subsection{Rare decays}

In the SM with massless $\nu$'s,
lepton-flavour-violating decays
(such as $D^0\to e^\pm\mu^\mp$, $X e^\pm\mu^\mp$) are completely
forbidden.
Flavour-changing neutral-current decays (such
as $D^0\to l^+ l^-$, $X l^+ l^-$, $X\nu\bar\nu$ and $X\gamma$) may occur in
the SM but they are highly suppressed by the GIM mechanism.
Rare $D$ decays are then very sensitive to
new physics, which may come from contact  interactions, leptoquarks,
horizontal gauge bosons, substructure, new scalars, technicolor,
etc. \cite{willey}.
Searches for rare decays of $D$ mesons are complementary to those
of $K$ or $B$ mesons because the new physics may be
flavour-dependent, i.e. different for up-like and down-like quarks.
Moreover,
the \tcf\ is probably the only machine that could measure
some radiative decays, such as $D^0\to\bar K^{*0}\gamma$,
at the low levels expected in the SM.

Tagging and precise beam-constrained mass measurements suppress
backgrounds to these processes. The \tcf\ would be sensitive to branching
ratios of ${\cal O}(10^{-8})$ \cite{schindler,stockdale}.
Present limits are in the range of $10^{-3}$ to $10^{-5}$.
Thus an improvement of more than 3 orders of magnitude is expected.

\subsection{$D^0$--$\bar D^0$ mixing}

Meson mixing has only been observed so far in the $K^0$--$\bar K^0$ and
$B^0$--$\bar B^0$ systems, where the second-order
$\Delta F = 2$ ($F = S, B$) transition is associated with a quark of charge
$-1/3$ (s, b). The study of mixing in the $D^0$--$\bar D^0$ system,
which contains a quark of charge $+2/3$, is then a fundamental experiment
for the understanding of the flavour structure of the weak interaction.
The rate for $D^0$--$\bar D^0$ oscillations is expected to be quite
small in the SM \cite{bigi_slac},
\be
r_D \equiv {{\rm Br}(D^0\to\bar D^0\to\bar f) \over
{\rm Br}(D^0\to f)} \sim 10^{-5} - 10^{-4} .
\ee
The reasons for such a suppression (GIM mechanism)
are very specific to the structure of the SM
and so many of its extensions (SUSY models, vector-like
fermions, E6-like models, etc.) lead to enhanced transition amplitudes.
$D^0$--$\bar D^0$ mixing is an ideal place for non-standard
flavour-changing neutral-currents to show up.

Signatures of mixing are like-sign dileptons from dual semileptonic
decays ($D^0\bar D^0\to l^\pm l^\pm X$) or
dual identical hadronic decays, such as
$D^0\bar D^0\to (K^\pm\pi^\mp)(K^\pm\pi^\mp)$.
At the \tcf,
the latter can be
separated from doubly Cabibbo-suppressed decays since quantum statistics
yield different correlations in the $D\bar D$ decays from $D^0\bar D^0,
D^0\bar D^0\gamma$, and $D^0\bar D^0\pi^0$ \cite{bigi_slac}.
When the $D^0$ mesons are in a relative P-wave, Bose statistics forbids
them to
decay into the same final state without mixing. In an S-wave, however, the
same final state can be reached through both, mixing and doubly
Cabibbo-suppressed decay amplitudes. By selectively running
at different c.m. energies, preparing well defined initial state conditions,
the mixing signature can be isolated from doubly-Cabibbo suppressed decays
or from new physics involving a violation of the $\Delta C = - \Delta Q_l$
($Q_l$ denotes leptonic charge)
selection rule.
Moreover, in the non-leptonic decay modes, the interference with the
doubly-Cabibbo suppressed
decay amplitude can be used to separate the mixing originating
in the mass ($x \equiv \Delta M_D/\Gamma_D$) and decay
($y\equiv \Delta\Gamma_D/ 2 \Gamma_D$) matrices ($r_D = (x^2+y^2)/2$)
\cite{bigi_slac}.

One can  also search for mixing in final states that do not
involve only neutral
$D$'s. For example one can look \cite{seiden,gladding}
at $D^+ D^{*-}$ final states, with a decay
chain:
$\; D^+\to K^-\pi^+\pi^+, \; D^{*-}\to\pi^- D^0\to\pi^- \, + \, l^\pm X$.
These decays have a very nice clean signal because
there is only one missing neutrino.

The \tcf\
could observe mixing at the limit  $r_D \approx 2\times
10^{-5}$ with one year's data \cite{schindler,gladding,fry},
to be compared with
the present 90\% C.L. limit,  $r_D < 3.7 \times 10^{-3}$.
Thus one should be able to bring the limit down to where one can expect to see
an effect even in the SM.

\subsection{CP violation}

In the three-generation SM, CP violation originates from the single
phase naturally occurring in the CKM quark-mixing matrix.
The present experimental observations are in agreement with
the SM expectations; nevertheless, the correctness of the CKM
mechanism  is far from being proved.
Like fermion masses and quark-mixing angles, the origin of the
CKM phase lies in the more obscure part of the SM Lagrangian:
the scalar sector. Obviously, CP violation could well be
a sensitive probe for new physics beyond the SM.

  A rate difference between $D^0 \bar D^0 \to (l^+ X)(f)$ and
$\bar D^0 D^0 \to (l^- X)(f)$, with $f$ a CP eigenstate (e.g.
$K^+K^-$) would provide an unambigous signal of CP violation, either
direct or involving $D^0 \bar D^0$ mixing.
Again, one can exploit quantum statistics \cite{bigi_slac} to disentangle
both effects.
Comparing possible CP asymmetries that can emerge in the reactions
$e^+e^-\to D^0 \bar D^0$,
$e^+ e^- \to D^0 D^{0*} + {\rm h.c.} \to D^0 \bar D^0\gamma$ and
$e^+ e^- \to D^0 D^{0*} + {\rm h.c.} \to D^0 \bar D^0\pi^0$,
it is possible to analyze the origin of the signal. In the absence of direct
CP-violation, only the second process, where the $D^0$--$\bar D^0$ system
is produced in a $C=+$ configuration, could generate a rate asymmetry.
Any asymmetry appearing in the other two processes should therefore be due
either to direct CP-violation in the decay amplitude or to detector bias.
A CP asymmetry of about 1\% could be measured \cite{gladding,fry} in
a one-year sample of $D^0 \bar D^0 \gamma$ events at E$_{cm}$ = 4.14 GeV.

CP violation can also be searched for directly in the rate
 in the process $\psi''(3.77)\to D^0 \bar D^0\to
f(D^0) f(\bar D^0)$, with $f$'s of the same CP parity, such as $f = K^+K^-$.
A single event of this type would establish  CP violation, since the
initial state is CP-even whereas the final state (P-wave) is CP-odd.
Although this method is not competitive with the asymmetry measurement for
determining the magnitude of the CP-violating amplitude (the rate
is proportional
to the square of the amplitude, while asymmetries have a linear dependence),
it is important in that it is sensitive both to the magnitude and
to the quantum mechanical phase
of the decay amplitudes \cite{bigi_slac}.
One expects to achieve $10^{-2}$ to $10^{-3}$ sensitivity
at the \tcf, if adequate background rejection is attained
\cite{seiden,gladding}.

In the SM, indirect CP violation from the
$D^0$--$\bar D^0$ mixing is expected to be very small.
Much better looks the possibility of detecting direct
CP-violating signals.
Recent calculations \cite{pugliese} for charged $D$ mesons
indicate CP asymmetries of the order of
$10^{-3}$ or larger for several final states, and other estimates
\cite{bigi_slac}
suggest that direct CP violation from neutral $D$ mesons should be
of a similar size.
The $D^+$ decay modes
$D^+\to K^+\bar K^{*0}\to K^+ K^-\pi^+$ and
$D^+\to\pi^+\eta\to\pi^+\gamma\gamma$, which are easy to reconstruct
and have small backgrounds, look quite promising;
a sensitivity of about $2\times 10^{-3}$
for the corresponding CP asymmetries,
i.e. at the level of the SM prediction \cite{pugliese},
could be achieved in a one-year data sample \cite{fry}.
Assuming that the indirect CP-violation amplitude is small,
a similar sensitivity could be achieved in several $D^0/\bar D^0$
decay modes
like $K^+K^-$, $\pi^+\pi^-$, $2\pi^+ 2\pi^-$, \dots
\cite{fry}.

The \tcf\ would
provide an important first window on CP violation in the up-quark sector.
This opens up the exciting possibility of exploring the mechanism
of CP violation in a system that is complementary to the $B$ or $K$ systems
and, moreover, sensitive to sources beyond the SM.
The measurement of direct CP violation at the level currently
predicted by the SM is clearly feasible at the \tcf,
although it will take several years of running to give definite results
\cite{fry}.

\subsection{Non-leptonic decays}

Non-leptonic charm decays provide a unique opportunity to probe
and understand the strong interactions in the interface between
the short-distance and long-distance regime.
Non-perturbative effects are much larger in $c$ decays than in
$b$ decays; but they are not overwhelming as in $s$ decays.
Thus, QCD corrections which need to be understood with high
precision in the beauty system,
are quantitatively enhanced in charm decays, allowing for
a very detailed analysis to be made.
Charm decays serve as Nature's microscope magnifying the
non-perturbative corrections affecting beauty decays \cite{bigi}.

Modern theoretical technologies (heavy-quark expansions, lattice,
QCD sum rules, $1/N_c$ expansion, \ldots)
are reaching a level where a treatment of charm (and beauty)
decays genuinely
rooted on QCD begins to be possible \cite{shifman}.
The rich variety of available charm decay modes
(meson and baryon decays; Cabibbo allowed, Cabibbo suppressed,
doubly Cabibbo suppressed)
offers an ideal laboratory to exhaustively test the
different theoretical methods, making possible a
reliable extrapolation to beauty decays.

The present experimental precision on charm decay properties
\cite{roudeau,weinstein2}
is certainly not satisfactory.
The largest part of $D^0$ and $D^+$ decay modes have an accuracy
worse than 20\%, while for the $D^+_s$  even
the absolute scale of its decay branching fractions
is not known in a reliable way.
Very few of the charm-baryon decay modes have been measured so far.
In many cases, a badly measured charm-decay branching ratio
is the bottleneck
to precise $b$ measurements at the BF or at LEP.
Fragmentation function studies would  greatly benefit from
a much better experimental knowledge of charm decays.
An accurate data basis on charm branching ratios would also allow us to
refine the predictions on CP-violating asymmetries in charm decays.

The \tcf\ is ideally suited to perform a  comprehensive
analysis of charm decays, including  $D^0$, $D^+$ $D^+_s$,
$\Lambda_c$, $\Xi_c^+$, $\Xi^0_c$ and $\Omega_c$
(c.m. energies of about 5.6 GeV have to be reached to look for
the heavier baryons).
A \tcf\ provides the only way --through the unique
capability to single-tag each of the various charm hadrons--
to improve the precision of the absolute branching ratios
for $D$ mesons to the per cent level, and to establish
absolute branching ratios at the 5\% level for $D_s$,
$\Lambda_c$, $\Xi_c$, etc.
The high-luminosity of the \tcf\ would turn out to be
very useful, since separate runs at different energy settings
are clearly needed.
Charm $P$ states could also be easily studied,
providing for the first time an absolute measurement
of their strong and electromagnetic decay branching ratios \cite{roudeau}.

\section{CHARMONIUM PHYSICS AND LIGHT-HADRON DYNAMICS}

\subsection{Charmonium and gluonium spectroscopies}

The charmonium system is an important source of information
for QCD studies and hadron spectroscopy.
The \tcf\ would provide a huge increase of the existing world data
sample and therefore would have a major impact on this area of physics.
In one year, the \tcf\ could produce $1.7\times 10^{10}$ $J/\psi$
events, i.e. three orders of magnitude more than the presently
accumulated statistics. The radiative decay
$J/\psi\to\eta_c(1S)\gamma$ would then generate a sample of $2\times 10^8$
$\eta_c(1S)$'s. Running at 3.686 GeV, $4\times 10^9$ $\psi'(3.69)$'s
would be produced in one year, giving rise to the secondary
production of $10^9$ tagged  $J/\psi$'s (through
$\psi'\to J/\psi\pi^+\pi^-$),
$3\times 10^8$ events of each type ($J=0,1,2$) of $\chi_{cJ}$ (via
$\psi'\to\gamma\chi_{cJ}$ decays),
and $10^7$ $\eta_c(1S)$'s and $\eta_c(2S)$'s (through the radiative
decays $\psi'\to\eta_c\gamma$).
Thus, the \tcf\ can in fact be considered as a charmonium factory.
Note that these high rates imply a huge number of secondary light mesons;
for instance, the decays $J/\psi\to\eta\gamma , \eta'\gamma$ would
give rise to more than $10^7$ tagged $\eta$ and $\eta'$ mesons,
largely improving the existing data samples \cite{benayoun}.

The $c \bar c$ wave function could be studied
via $J/\psi\to 3 \gamma$ and $\eta_c\to 2 \gamma$, which constitute direct
tests of the charmonium models.
Presently open questions on the charmonium electromagnetic couplings
could be answered through accurate measurements at the \tcf\
\cite{toki,barnes,seth,franzini}.
Especially interesting
are the M1 decays $J/\psi\to\gamma\eta_c$ and
$\psi'\to\gamma\eta_c,\gamma\eta'_c$, which could discriminate between a
wide range of theoretical predictions \cite{barnes},
and the E1 transition $\chi_{c2}\to J/\psi\gamma$,
which can be used to study a possible anomalous magnetic moment of
the $c$ quark, by measuring the photon angular distribution
\cite{barnes,seth}.
The \tcf\ detector should be ideally suited for these measurements because
of its extremely low threshold ($E_{\mbox{\rms min}}^\gamma = 10$ MeV)
and its $\geq$ 99\% hermeticity \cite{seth}.
Improved measurements on radiative and hadronic charmonium transitions
would allow to test our theoretical understanding of quarkonium dynamics
\cite{dibartolomeo2}.
The huge amount of $\eta_c$
and $\chi$ events  could be used to study many hadronic
decays of these states \cite{mir}.

The \tcf\ could also complete
the charmonium spectroscopy with the confirmation  of the
 $^1P_1(3526)$ and $\eta^{\prime}_c(3590)$ states \cite{seth},
and the discovery of the missing $^3D_2$ and $^1D_2$ states.
Their quantum numbers can be accessed at the \tcf\ through
the decays: $\psi^*\to\eta\, ^1P_1$ ($\sqrt{s}> 4$ GeV),
$\psi'\to\gamma\eta'_c$, $\psi^*\to\eta\, ^3D_2$
($\sqrt{s}> 4.5$ GeV) and $\psi^*\to\eta\, ^1D_2$
($\sqrt{s}= 4.03$ GeV) \cite{close}.

Excited $\psi^*$ states above the $D\bar D$ threshold are
very badly known at present.
It would be important to measure their decay branching fractions
to hadronic final states.
Measurements of
the relative fractions of $D$, $D^*$ and
$D^{**}$ in their decays would provide important information and tests
of Heavy Quark Effective Theory, and could reveal the internal
structure of these states \cite{barnes,close}.

The \tcf\ would make a systematic search for gluonia and hybrids in
a gluon-rich  environment: $J/\psi\to \gamma g g,\; ggg$; $\eta_c\to
gg$.   For example, the decay $J/\psi\to\gamma g g \to \gamma X$ involves a
pure two-gluon intermediate state with a mass $m_{gg}\leq 3.1$ GeV,  i.e. in
the expected mass region of the gluonium spectrum.  Furthermore,  the
nature of the state $X$ can be tested experimentally by comparing
$J/\psi\to\gamma X, \omega X$ and $\phi X$ \cite{doser}.
 Resonances that appear in
radiative decays but are suppressed with $\omega$ and $\phi$ are likely to
be gluonia.
The comparison with the two-photon production process\footnote{
It is worth mentioning that the presence in the \tcf\ detector
of an electromagnetic
calorimeter at small angles  allows the detection
of single and double tag events in two-photon physics.
This is a unique possibility among the existing machines
\protect\cite{palano}.}
$\gamma\gamma\to X$ can further help to discriminate between
$q\bar q$ states and gluonium candidates \cite{close,palano}:
since gluons do not carry electric charge, glueballs should not
couple to $\gamma$'s.
In addition,
$J/\psi\to\gamma X\to\gamma (\gamma\rho:\gamma\omega:\gamma\phi)$
can be used to filter the flavour content of all $C=+$ states;
moreover, for $J_X\geq 2$ the relative helicity amplitudes for
$X\to\gamma V$ may also be measured and give
information on the internal structure of $X$ \cite{close}.
This requires the \tcf\ statistics since only the
channel $J/\psi\to\gamma\gamma\rho$ has been measurable so far.

At the $\psi'$ peak, the huge number of produced $\chi_{cJ}$'s
can be used to constrain the partial wave analysis
of hadronic decays, by fixing the spin $J$ of the initial $\chi_c$.
For instance, $\chi_{c1}\to\pi H$ is sensitive to the
hybrid exotic sector $H$ ($J^{PC}=1^{-+}$),
while
$\chi_{c0}\to f_0(975) X$ provides a gateway to  $0^{++}$ hadrons
\cite{close}.
At present there is basically no data on the $\chi_{cJ}$ states.
With the statistics of the \tcf\, the physics output
from $\chi_{cJ}$ decays could exceed that already known from the
$J/\psi$.

Charm hybrids $H_c$ are expected to occur  around the
$DD^*$ threshold. These hybrids include vector states which can be
formed directly in $e^+e^-$ annihilation. Moreover,
theoretical considerations suggest that charm is the optimal flavour
to unambiguously identify vector hybrid states: they are light
enough to be produced in $e^+e^-$ annihilation with leptonic
widths of $\cO(0.1)$ keV, and heavy enough that the
conventional quarkonium states are well understood so that extra
states can be readily identified \cite{close}.
A fine grained energy scan to search for these states could be
carried out at the \tcf\ within two weeks; this involves
measuring the hadronic cross-section above charm threshold in 1 MeV
steps to an accuracy of 2\% \cite{close}.
The implementation of a monochromator, enabling a very fine scan
with higher statistics would be very desirable to clarify the
nature of detailed structures.
Note, that a precise measurement of the hadronic cross-section in the
\tcf\ energy range would also provide important information for inclusive
tests of QCD; moreover, this information is needed to get accurate
predictions on  electroweak radiative corrections (e.g. $g-2$ or
LEP physics). The superb hermeticity and high efficiency
for hadron detection of the \tcf\ detector are ideal for this purpose
\cite{seth}.

Triggering on $J/\psi$'s at any operating energy of the \tcf\ may
reveal the existence of other exotic states that are not
directly produced in $e^+e^-$ annihilation \cite{close}.
For instance, it may be possible to detect $\psi_H$ hybrids
through $e^+e^-\to\eta\psi_H\to\eta(\eta J/\psi)$.

The existence of glueballs, hybrids and other exotics,
is a natural consequence of QCD. Unfortunately these states have
proved to be very elusive and none of them has been
experimentally established so far.
The main difficulty comes from the absence of clear-cut signatures
to distinguish them from the usual $q\bar q$ states.
The basic criterion is then the observation of an extra state,
which has no room in a $q\bar q$ nonet.
One needs detailed knowledge about their production mechanisms
and decay patterns, besides measuring their quantum numbers.
With its very high statistics and low backgrounds,
the \tcf\ is certainly the best machine to perform such an
exhaustive analysis.
The unambiguous identification of a glueball or an hybrid state
would be a fundamental discovery.

\subsection{Weak decays and axions}

The very high statistics and the
OZI suppression of the $J/\psi$ decays, causing a
narrow width,
may provide the first opportunity to measure  weak decays
of a vector meson \cite{toki,sanchis}, such as ($l=e,\mu$)
$J/\psi\to D_s l \nu_l$ [Br $\simeq 10^{-8}$], or the C-violating decay
$J/\psi\to\phi\phi$ [Br $\simeq 10^{-8}$].

The $J/\psi$ can be tagged via
$\psi'\to\pi^+\pi^- J/\psi$ to allow searches for rare processes
\cite{toki} such as the
neutral-current decay $J/\psi\to\nu\bar\nu$, which in the
SM is expected  at  a very low branching ratio of $N_\nu\times 10^{-8}$;
if any signal is seen it would be a clear indication of new physics.
The limiting background would be $\jpsi\to n \bar n$, which has a
branching ratio of 0.18\% \cite{toki}.

Axions or other evasive neutrals can be produced in
$J/\psi\to\gamma X$. The signature for such a decay is a single photon in
the final state. The limiting factor here will be backgrounds from
events with 3 photons in which only one photon is detected. Assuming the
photon detection efficiency is 99\%, the expected sensitivity is
Br($J/\psi\to\gamma X) \simeq 10^{-8}$.

\subsection{CP violation}

At the $\jpsi$ peak, a huge number of hyperon pairs would be
produced through the decay $\jpsi\to Y \,\bar Y$ ($Y = \Lambda, \Xi$).
The $1.7\times 10^{10}$ $\jpsi$'s accumulated in one-year's data,
would give rise to
$2.3\times 10^7$ $\Lambda \,\bar\Lambda$ and
$3.1\times 10^7$ $\Xi\,\bar\Xi$ pairs.
These numbers could be substantially increased
with the implementation of a monochromator
(a narrower beam spread implies a much larger visible
$\jpsi$ production cross-section at the peak).
With 100 keV monochromator optics,
the \tcf\ could produce
$8\times 10^{10}$ $\jpsi$ events in one year
and, therefore,
$1.1\times 10^8$ $\Lambda \,\bar\Lambda$ and
$1.4\times 10^8$ $\Xi\,\bar\Xi$ pairs
would be generated.
A high-precision study of hyperon-decay properties would obviously
become possible. Moreover,
this hyperon factory could be used to search for signals of
direct CP violation \cite{gonzalez}.

A complete description of the decay $Y\to B\pi$ involves,
in addition to the rate $\Gamma_Y$, two decay parameters
usually denoted $\alpha_Y$ and $\beta_Y$.
$\alpha_Y$ regulates the angular distribution of the final baryon
in the decay of a polarized hyperon (in the $Y$ rest frame),
while $\beta_Y$ characterizes the
$B$ polarization in the direction normal to the plane defined by
its momentum and the spin of the decaying hyperon.
CP invariance requires $\Gamma_{\bar Y} = \Gamma_Y$,
$\alpha_{\bar Y} = -\alpha_Y$ and $\beta_{\bar Y} = - \beta_Y$.
Thus, one can consider the following CP-violating asymmetries:
\be
\Delta_Y \equiv {\Gamma_Y - \Gamma_{\bar Y} \over
\Gamma_Y + \Gamma_{\bar Y}} ,
\qquad\quad
A_Y \equiv  {\alpha_Y + \alpha_{\bar Y}\over
\alpha_Y - \alpha_{\bar Y}} ,
\qquad\quad
B'_Y \equiv {\beta_Y + \beta_{\bar Y} \over \alpha_Y - \alpha_{\bar Y}} .
\ee
The SM predictions for these asymmetries are quite
uncertain, due to the important role of non-perturbative QCD effects
in the decay amplitudes.
In a model-independent fashion, the relative size of the different signals
is expected to be \cite{steger}:
\be
B'_Y \,\approx\, 10 \, A_Y \,\approx\, 100 \,\Delta_Y .
\ee
A recent update of several model-dependent calculations \cite{steger}
quotes values of $A_Y$ between $10^{-5}$ and $10^{-4}$.
To get better theoretical estimates, it is necessary to
experimentally improve the knowledge of the
(CP-conserving) decay amplitudes
and the final strong-interaction phases.

Although $\Delta_Y$ is the easiest to measure, it has the
lowest sensitivity; on the other side,
a determination of $B'_Y$ is difficult since it requires
information on the spins of both baryons.
Therefore, the experiments have focussed mainly in
measuring $A_Y$.
The best results have been obtained by PS185 (LEAR)
with $1.5 \times 10^4$ $\Lambda\to p\pi^-$ events, quoting
$A_\Lambda = -0.024 \pm 0.057$ \cite{PS185}.


In $e^+e^-\to\jpsi\to Y \,\bar Y$, the hyperons are produced unpolarized
(assuming parity conservation in the $Y\,\bar Y$ production);
nevertheless, their spins are strongly correlated.
The information on the spin-dependent part
of the production amplitude can then be reconstructed from the
correlated distribution of their decay products.
Analyzing the process
\be
\jpsi\to\Lambda\;\bar\Lambda \to
  (p\pi^-) \; (\bar{p}\pi^+) ,
\ee
one can determine the value of $\alpha_\Lambda \alpha_{\bar\Lambda}$
(the spin correlations obviously involve products of both spins).
Using the value of $\alpha_\Lambda$ measured somewhere else \cite{PDG92}
the asymmetry $A_\Lambda$ is then obtained.
A direct separate measurement of $\alpha_\Lambda$ and
$\alpha_{\bar\Lambda}$ would require polarized $\Lambda$'s;
this could be possible with polarized $e^+e^-$ beams.

The decay chain
\be
\jpsi\to\Xi^-\;\bar\Xi^+\to (\Lambda\pi^-)\;(\bar\Lambda\pi^+) \to
  (p\pi^-\pi^-) \; (\bar{p}\pi^+\pi^+) ,
\ee
allows to extract a much richer information. The $\Lambda\,\bar\Lambda$
correlation measures $\alpha_\Xi\alpha_{\bar\Xi}$, as before,
but now $\Lambda$'s are polarized and this polarization is analyzed
by the final proton distributions. Making a global analysis of
all correlated distributions the separate values of
$\alpha_\Xi$, $\alpha_{\bar\Xi}$, $\beta_\Xi$, $\beta_{\bar\Xi}$,
$\alpha_\Lambda$ and $\alpha_{\bar\Lambda}$ can be measured.
Therefore, one obtains a determination of the asymmetries
$A_\Xi$, $B'_\Xi$ and $A_\Lambda$.
The access to $B'_Y$ is a welcome feature, since this asymmetry
has the highest sensitivity.

\begin{table}[htb]
\caption{Sensitivities \protect\cite{gonzalez}
   to the different CP-odd $\jpsi\to Y\,\bar Y$
   observables at the \tcf, with
   $8\times 10^{10}$ $\jpsi$ events (1 year-data with monochromator).
   Estimates for a possible 40\% beam polarization are also given.}
\label{tab:hyperons}
\begin{center}
\begin{tabular}{|c|c|c|c|}
\hline
Decay Mode & Beam Polarization &  Measured Observable & \tcf sensitivity
\\ \hline
$\Lambda\,\bar\Lambda$ & No & $A_\Lambda$ & $5\times 10^{-3}$
\\ \hline
$\Lambda\,\bar\Lambda$ & Yes & $A_\Lambda$ & $\leq 3\times 10^{-3}$
\\ \hline
&& $A_\Lambda$ & $10^{-3}$
\\
$\Xi^-\Xi^+$ & No & $A_\Xi$ & $10^{-3}$
\\
&& $B'_\Xi$ & $5\times 10^{-3}$
\\ \hline
&& $A_\Lambda$ & $\leq 10^{-3}$
\\
$\Xi^-\Xi^+$ & Yes & $A_\Xi$ & $\leq 10^{-3}$
\\
&& $B'_\Xi$ & $\leq 5\times 10^{-3}$
\\ \hline
\end{tabular}
\end{center}
\end{table}

Table \ref{tab:hyperons} \cite{gonzalez}
shows the expected sensitivity at the \tcf,
with $8\times 10^{10}$ $\jpsi$ events (1 year-data with monochromator).
Although still far away from the expected SM signal,
the $A_Y$ measurement would represent an improvement better than one order
of magnitude over present data.
More significative is the estimated accuracy attainable for
the asymmetry $B'_\Xi$, reaching sensitivities that are
within the values expected in the SM.
The big uncertainties of the theoretical predictions
would make difficult to extract useful information on the CKM phase
from this measurement.
Nevertheless, the experimental observation of a non-zero
CP-violating asymmetry would be a major milestone in our
understanding of CP violation, as it would clearly establish the
existence of direct CP violation in the decay amplitudes.

\section{SUMMARY}


The flavour structure of the SM is one of the main pending questions
in our understanding of weak interactions.
Although we do not know the reason of the observed
family replication, we have learn experimentally that the
number of SM fermion generations is just three (and no more).
Therefore, we must study as precisely as possible
the few existing flavours,
to get some hints on the dynamics responsible for their
observed structure.
The construction of high-precision
flavour factories is clearly needed.

Without any doubt, the \tcf\ is the best available tool to explore
the $\tau$ and $\nu_\tau$ leptons and the charm quark.
This facility combines the three ingredients required
for making an accurate and exhaustive investigation of these
particles:
high statistics, low backgrounds and good control of systematic errors.
The threshold region provides a series of unique features
(low and measurable backgrounds free from heavy flavour contaminations,
monochomatic particles from two-body decays, small radiative corrections,
single tagging, high-rate calibration sources, \ldots)
that create an ideal experimental environment for this physics.

As shown in the previous sections, the \tcf\ would cover
an exceptionally broad programme of solid physics.
The SM of electroweak interactions would be accurately proved in
$\tau$, charm and charmonium decays
(universality tests, Lorentz structure of weak interactions, neutrino
masses, flavour mixing, CP violation, $D^0$--$\bar D^0$ oscillations,
lepton-flavour violation,
flavour-changing neutral currents, rare decays, \dots).
Moreover, the \tcf\ would constitute a superb
laboratory to test QCD at the interface of perturbative and
non-perturbative dynamics
(the $\tau$ is the only lepton with hadronic decays,
important interplay of strong interactions in charm decays,
light-hadron and gluonium spectroscopy). In fact,
as emphasized by several speakers at this workshop, the \tcf\
is optimally suited to be considered as ``the QCD machine for
the 90's''.

One can identify a series of different energies
that are optimized for specific physics measurements
(nevertheless, different studies can share the same energy setting;
for instance, $\tau$ analyses below the charm threshold
allow to measure the backgrounds for charm measurements).
The high-luminosity of the \tcf\ is a crucial
requirement to efficiently operate the machine at different points.
Of course, for specific worth-while investigations (or once
an indication for a major discovery surfaces),
the high luminosity can be concentrated at the appropriate energy.
In any case, it is clear that the \tcf\ is a facility
for performing a long-term research programme.

In conclusion, the \tcf\ could address fundamental aspects of the SM
that are complementary to those addressed by higher-energy machines.
A comprehensive set of precision measurements for $\tau$, charm
and light-hadron spectroscopy would be obtained, proving the SM
to a much deeper level of sensitivity and
exploring the frontier of its possible extensions.

\section*{Acknowledgements}

I would like to thank the organizers for creating
a very stimulating atmosphere and  all the
participants of this workshop for their work and ideas,
which I have extensively used to prepare this summary.

\end{document}